\documentclass{elsart}
\usepackage{graphicx}
\usepackage{graphics}
\usepackage{amssymb}
\usepackage{amsmath}
\usepackage{bm}
\usepackage{slashed}
\usepackage{lineno}
\newcommand{\g}{\gamma}
\newcommand{\gmu}{\gamma^\mu}
\newcommand{\gnu}{\gamma^\nu}

\newcommand{\gmunu} {g^{\mu\nu}}
\newcommand{\kmu}{k^\mu}

\newcommand{\kbnu}{\bar{k}^\nu}

\newcommand{\pmu}{p^\mu}

\newcommand{\pbnu}{\bar{p}^\nu}

\newcommand{\Nmu}{N^\mu}

\newcommand{\nmu}{n^\mu}

\newcommand{\smu}{s^\mu}
\newcommand{\snu}{s^\nu}

\newcommand{\be}{\begin{equation}}
\newcommand{\ee}{\end{equation}}
\newcommand{\bea}{\begin{eqnarray}}
\newcommand{\eea}{\end{eqnarray}}
\newcommand{\beal}{\begin{align}}
\newcommand{\eal}{\end{align}}
\newcommand{\bespl}{\begin{split}}
\newcommand{\espl}{\end{split}}

\newcommand{\nsl}{\kern 0.2 em n\kern -0.50em /}
\newcommand{\ksl}{\kern 0.2 em k\kern -0.45em /}
\newcommand{\psl}{\kern 0.2 em p\kern -0.50em /}
\newcommand{\Nsl}{\kern 0.2 em N\kern -0.50em /}
\newcommand{\ssl}{\kern 0.2 em s\kern -0.50em /}
\newcommand{\pbsl}{\kern 0.2 em \bar{p}\kern -0.50em /}
\newcommand{\sbsl}{\kern 0.2 em \bar{s}\kern -0.50em /}
\newcommand{\kbsl}{\kern 0.2 em \bar{k}\kern -0.50em /}
\newcommand{\nbsl}{\kern 0.2 em \bar{n}\kern -0.50em /}
\newcommand{\Nbsl}{\kern 0.2 em \bar{N}\kern -0.50em /}
\newcommand{\Pslash}{\kern 0.2 em P\kern -0.50em /}
\newcommand{\Rslash}{\kern 0.2 em R\kern -0.50em /}

\begin{document}

\begin{frontmatter}

\title{Inserting physics associated with the transverse polarization 
of the quarks into a standard Monte Carlo generator,  
without touching the code itself. 
} 

\author{A.~Bianconi} 
\address{Dipartimento di Chimica e Fisica per l'Ingegneria e per i 
Materiali, Universit\`a di Brescia, I-25123 Brescia, Italy, and\\
Istituto Nazionale di Fisica Nucleare, Gruppo di Brescia, 
I-25100 Brescia, Italy.}
\ead{Email: bianconi@bs.infn.it}

\begin{abstract}
The transverse polarization of a quark is a degree of freedom that is 
not taken into account in the most commonly used Monte Carlo generators. 
For the case $e^+e^-$ $\rightarrow$ $hadrons$ I show that it is possible 
to use these generators to simulate processes where 
the parent quark and antiquark are transversely polarized and the 
fragmentation process is affected by this polarization. The key point 
is that it is possible to obtain this without touching the generator 
code at all. One only works on the parton-level and hadron-level outputs 
that the Monte Carlo code has produced, modifying them in a correlated way. 
A group of techniques is 
presented to obtain this, matching the most obvious needs of a user 
(in particular: reproducing a pre-assigned final distribution). 
As an example these methods are applied to modify Pythia-generated events 
to obtain a nonzero Collins function and a 
consequent $cos(\Phi_1+\Phi_2)-$asymmetry 
of pion pairs. 
\end{abstract}


\end{frontmatter}

\maketitle

\section{Introduction}

A great effort covering several decades has been devoted to 
developing full-purpose MonteCarlo codes (e.g. 
\cite{Pythia8,Lepto,Herwigpp}, see \cite{Buckley11} for a more general 
review) for high-energy hard processes. 
However, the most known codes do not allow for simulation of effects like 
azimuthal asymmetries in SIDIS, in $e^+e^-$ $\rightarrow$ hadrons, in 
Drell-Yan processes etc, i.e. those processes where the 
transverse spin of the quark has a relevance (for an overview of 
this field, see the proceedings \cite{T08}, and the reviews \cite{BDR} 
and \cite{BBM}). 

Although it is possible to modify a Monte Carlo generator in such a way 
to produce such asymmetries,  
the perspective of touching the core of 
such complicate codes is surely a nightmare for most 
of the people that are potentially interested in. What I present here 
is a set of techniques aimed at introducing (transverse) spin physics in 
a generator without touching it at all. Of course one needs writing  
patches of code, but these use the parton level and hadron level 
outputs of the 
generator\footnote{ 
The most known MC generators show, in their output, 
the momenta of all the particles produced in the intermediate 
stages of a complex event. 
}
as an input, and are completely independent 
(small) programs. 

In this work, I will consider $e^+e^-$ $\rightarrow$ hadrons. 
I will
start from events generated by Pythia-8 \cite{Pythia8}. I need to modify 
Pythia outputs in such a way that: 

(i) In the hard vertex $e^+e^-$ $\rightarrow$ $q\bar{q}$,  
transverse polarizations are added to the quark and to the antiquark, 
and the correlated distribution of spins and 
momenta is coherent with the known matrix elements. 

(ii) The momenta of the final hadrons in each hemisphere 
(quark hemisphere and antiquark 
hemisphere) are modified in a way that is correlated with the 
transverse spin of the parent quark. The modification must be under 
our full control, i.e. we should be able to obtain exactly what we 
want to obtain. This may mean a fragmentation function with pre-assigned 
form, or the implementation of a model. 

I will describe some techniques to implement the previous 
requirements. As an example, I will apply 
them to modify pion final momenta so to have them distributed 
according with the sum of an unpolarized 
and a Collins fragmentation function\cite{CollinsFunction}. 
Although here I just want to show an example of application of 
these simulation techniques, the chosen case is of special 
interest, since the Collins function has a relevant 
role in the extraction of information on the transverse polarization 
of the nucleon\cite{Anselmino}, there are models 
for it \cite{Bacchetta1,Goldstein,Bacchetta2}, and it appears in 
asymmetries measured 
in SIDIS \cite{HermesCollins,CompassCollins,CompassCollins2} and 
$e+e^-$ $\rightarrow$ $hadrons$ \cite{BelleCollins1}. 

As a consequence of a nonzero Collins function, the 
correlated distribution of pions detected in opposite hemispheres 
is expected to present a $cos(\Phi_1+\Phi_2)$ asymmetry in the sum of the 
azimuthal angles of the pions \cite{BoerEpem1} (see section VIII of 
\cite{BoerBelle1} for details). 
This will be confirmed by an analysis of the events generated 
by Pythia and modified as suggested here. 

I will not try to reproduce 
the detailed physical outputs of the recent measurement of this quantity 
at Belle \cite{BelleCollins1,BelleCollins2}, because this is just an 
exercise and the aim of this work is to present a more general 
group of techniques. Once the individual hadron tracks have been 
modified in a physically motivated way, other 
azimuthal asymmetries could be generated in a set of simulated events, 
like the $cos(2\phi)-$asymmetry (see 
\cite{BoerBelle1} and \cite{BelleCollins2}), or 
dihadron-dihadron correlations \cite{RadiciDihadron1,BCR2011}, 
or even asymmetries 
associated with a larger number of particles\cite{EfremovHandedness}. 

Let me name 
NPMC the ``non polarized'' MonteCarlo code whose outputs have to be 
modified by the external patches. 

The two main steps, corresponding to previous (i) and (ii) are: 

Step (1): Take a $q\bar{q}$ pair produced in the hard vertex by the 
NPMC and ``stick'' a pair of reciprocally 
independent random transverse spins, in such a 
way that the correlated spin-momentum distribution agrees with the 
polarized quark-lepton squared matrix element. 
Transverse spins are 
assumed to be classical fixed-length vectors with one degree of freedom 
(the angle in the plane that is normal to the quark momentum).  

This step is the critical point of the method. 
It needs to be demonstrated that it is feasible. One thing is 
to sort both momenta and spins according with a joint distribution, and 
another thing is first sorting momenta according with the spin-averaged 
distribution (that is done by the NPMC), and next dividing the sorted 
events into fairly distributed spin subsets (that is done by us). 
Section 2 is devoted to this. 

Step (2): Modify the (final or intermediate) 
hadronic momenta so to reproduce the effect of an assigned 
quark-spin-dependent fragmentation function. Alternatively, one 
could like to implement a physical model that is behind this 
fragmentation function. 

Here the underlying assumption is that azimuthal effects 
are a $small$ 
distortion of a final particle distribution that is mainly determined 
by the physics implemented in the NPMC. 
There are two classes of techniques that may be exploited: 
``distortion'', and ``filtering'' techniques. 
In distortion techniques  
the individual particle properties in a given event are modified. 
These techniques exploit each event produced by the NPMC. 
In filtering techniques only a subset of the events produced 
by the NPMC is accepted. 



Filtering may be expensive, when many NPMC-events are needed to 
obtain one final event. It may be necessary when we think that the 
additional physical processes  
affect quantities like the final pion multiplicity, or the 
general structure of the event. If we do not expect this to be the case, 
distortion techniques are preferable. 

In section 3, I show the effect of two possible distortion 
techniques in producing a nonzero Collins function for the pions. 

A $cos(\Phi_1+\Phi_2)-$asymmetry 
involving pions of the opposite 
hemispheres is a synthesis of the presence of Collins 
functions on both sides, and of the spin correlations between the 
quark and antiquark produced in the hard vertex. So, if things in 
the steps (1) and (2) have been properly performed it must be 
present. Section 4 is devoted to show that this is the case, 
i.e. that a $cos(\Phi_1+\Phi_2)-$asymmetry is present in 
a set of Pythia-events 
modified by the presented techniques.

\section{Step 1: Hard vertex and spin}

In most MonteCarlo codes for high-energy physics, the starting 
point is the hard scattering process at parton level. Once 
a hard scattering event has been sorted 
according with some probability, both later 
and previous cascading processes are generated.

\begin{figure}[ht]
\centering
\includegraphics[width=9cm]{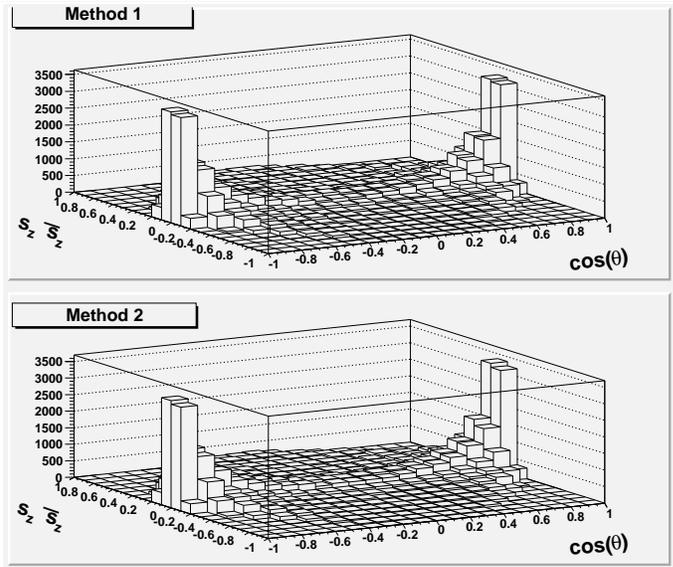}
\caption{
Event distribution as a function of the quark-lepton angle $\theta$ and 
of the 
components of spins of the quark and antiquark along the 
lepton axis. 
\label{fig:XX1}}
\end{figure}

\begin{figure}[ht]
\centering
\includegraphics[width=9cm]{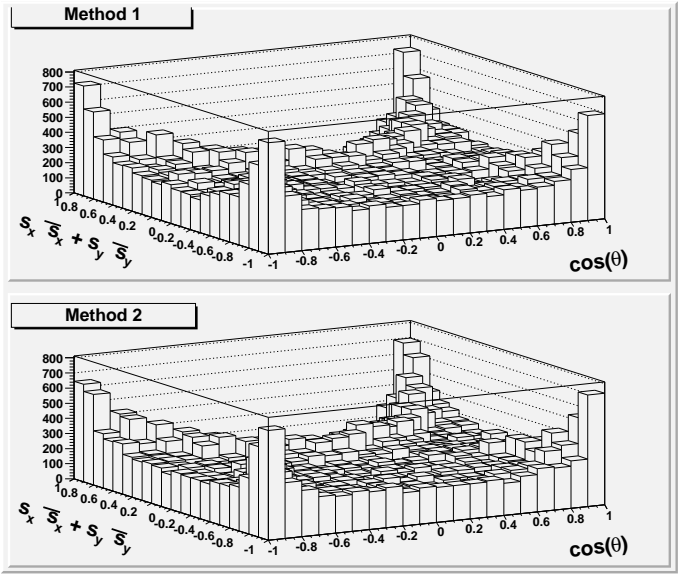}
\caption{
Event distribution as a function of the quark-lepton angle $\theta$ and of 
the 
components of spins of the quark and antiquark on the plane 
normal to the lepton axis (see the next figure for the individual 
contributions of $x$ and $y$ terms). 
\label{fig:XX2}}
\end{figure}

\begin{figure}[ht]
\centering
\includegraphics[width=9cm]{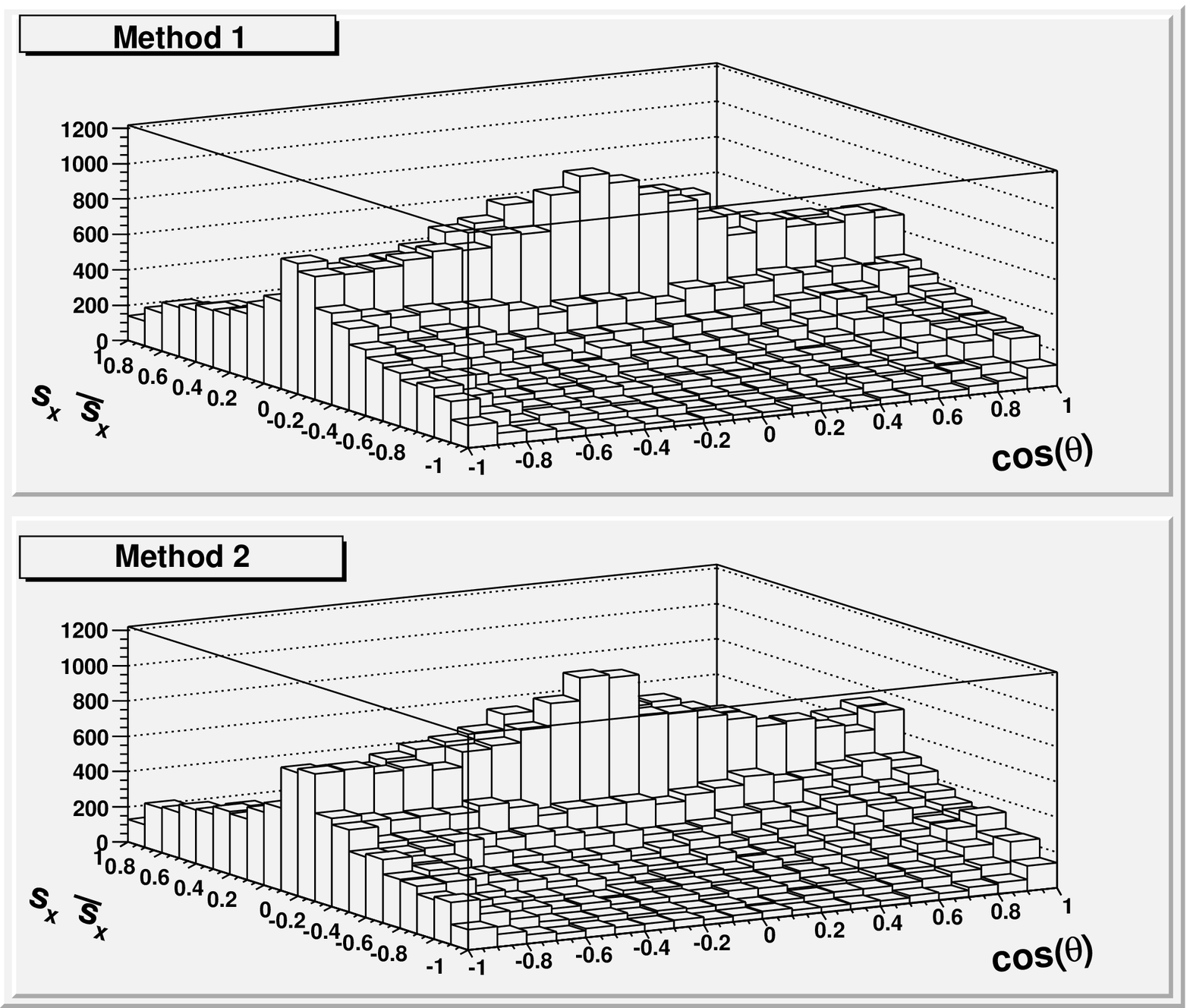}
\caption{
Event distribution as a function of the quark-lepton angle $\theta$ and of 
the 
components of spins of the quark and antiquark along the $x-$axis 
(the corresponding $y-$distribution is  equal). 
\label{fig:XX3}}
\end{figure}

\begin{figure}[ht]
\centering
\includegraphics[width=9cm]{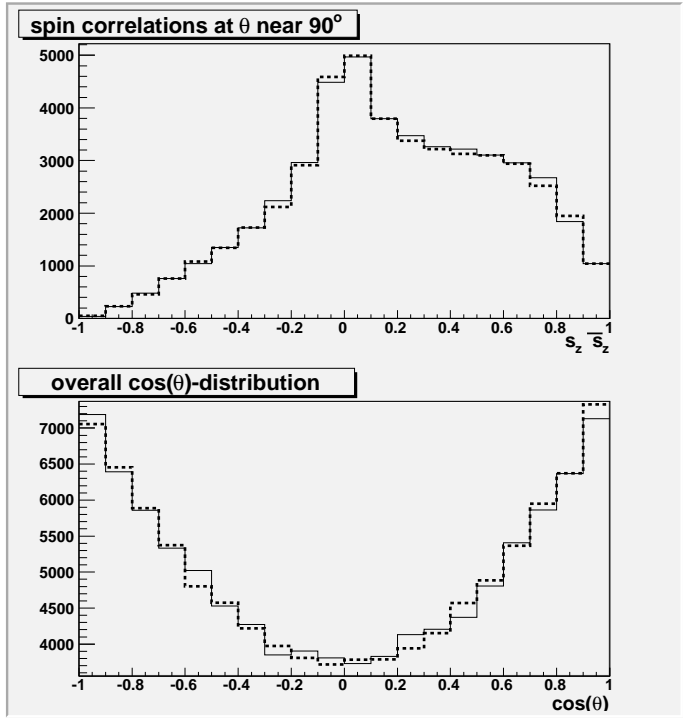}
\caption{
Upper panel: $s_z\bar{s}_z-$distribution integrated over the 
$cos(\theta)-$range $(-0.5, 0.5)$. Lower panel: 
the $cos(\theta)-$distribution, summed over spins. The continuous 
line is obtained with Method 1, the dotted line with Method 2.
\label{fig:XX4}}
\end{figure}

\begin{figure}[ht]
\centering
\includegraphics[width=9cm]{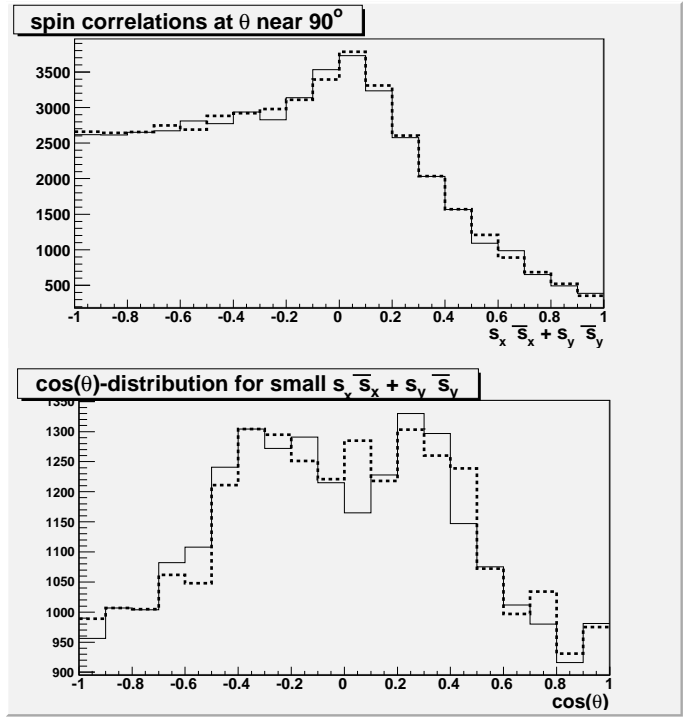}
\caption{
Upper panel: $s_x\bar{s}_x+s_y\bar{s}_y-$distribution integrated over the 
$cos(\theta)-$range $(-0.5, 0.5)$. Lower panel: the $cos(\theta-)$distribution, 
for $-0.2$ $<$ 
$s_x \bar{s}_x+s_y \bar{s}_y$ $<$ $0.2$. The continuous 
line is obtained with Method 1, the dotted line with Method 2.
\label{fig:XX5}}
\end{figure}

In the case of $e^+e^-$ $\rightarrow$ $hadrons$, the (parton level) 
cross section for the hard scattering 
process $e^+e^-$ $\rightarrow$ $q\bar{q}$ with transversely-polarized 
quarks, and with the leptons on the 
$z-$ axis, is (see Appendix): 

\noindent
\begin{eqnarray}
W_{pol}(\hat N, \vec s, \vec {\bar{s}})\ =\ 
{1 \over 4}
\Bigg(
[1\ +\ cos^2(\theta)] (1\ +\ s_z\bar{s}_z)\ 
\nonumber
\\
-\ sin^2(\theta) (s_x\bar{s}_x\ +\ s_y\bar{s}_y) 
\Bigg)
\label{eq:full_pol} 
\end{eqnarray}
where $\hat N$ is a versor of the quark momentum ($-\vec N$ 
is the corresponding one for the antiquark), 
$\theta$ 
is the angle between the lepton and quark axes 
($cos(\theta)$ $=$ $\hat N \cdot \hat z$) 
and $\vec s$, $\vec {\bar{s}}$ 
are transverse polarization vectors. ``Transverse''
means ``transverse to the (anti)quark 3-momenta''. Since 
the quark momentum is not aligned with the electron 
momentum, the transverse spin has a nonzero $s_z$ component, i.e. 
a component along the electron-positron beam axis. This must  
not be confused with the helicity or with the longitudinal spin 
of the quark. 

This cross section was already reported in \cite{AB_EPJA2} 
where it was applied to the Drell-Yan process in the context 
of the Panda experiment\cite{Panda}. 
The two cross sections 
present a different form because they are expressed in different 
reference frames, but they are the same. 

If the polarization components are averaged away, we get the known 
result 
\noindent
\begin{eqnarray}
W_{no pol}(\hat N)\ \propto\ 
1\ +\ cos^2(\theta).
\label{eq:full_nopol} 
\end{eqnarray}
The 
most known and used MonteCarlo codes implement this equation for the 
lepton-quark vertex, adding further $O(\alpha_s^n)$ hard processes in 
the partonic showers. 

If we had to write from the very beginning a 
complete MonteCarlo generator including transverse polarizations 
of the quark and antiquark, we could imagine two ways of sorting 
momenta $and$ spins of the quark and of the antiquark 
(generically: ``quarks''). 

Method 1) The momenta and the spins of the quarks are jointly sorted 
according with the probability law eq.\ref{eq:full_pol}. 

Method 2) First, the momenta of the quarks are sorted accordingly 
with eq.\ref{eq:full_nopol}, and next the polarizations are sorted in such 
a way to get the distribution \ref{eq:full_pol} for the joint set 
of variables $\hat N, \vec s, \vec {\bar{s}}$.  

The former one is the right one in general (if correctly implemented).  
The latter 
method allows for splitting the generation process into two steps, 
one of which may be performed by an NPMC and the other one by our 
additional code patches. I will show that in the case interesting us 
it produces the same results of method 1. 

To apply Method 2 here I take a quark-antiquark pair whose momenta have 
been sorted with probability $W_{no pol}(\hat N)$, and sort their 
spins according with the 
distribution 
\noindent
\begin{eqnarray}
D(\hat N, \vec s, \vec {\bar{s}})\ =\ 
{{W_{pol}(\hat N, \vec s, \vec {\bar{s}})}\over   {W_{no pol}(\hat N)} }
\ \rightarrow\ W_{pol}(\hat N, \vec s, \vec {\bar{s}})
\label{eq:Wratio} 
\end{eqnarray}
subject to the 
conditions 
\noindent
\begin{eqnarray}
\vec s \cdot \hat N =  \vec {\bar{s}} \cdot \hat N\ =\ 0, 
\nonumber 
\\
|\vec s|\ =\ 1,\  |\vec {\bar{s}}|\ =\ 1. 
\label{eq:orthogonality} 
\end{eqnarray}
In eq.\ref{eq:Wratio}, in general, 
$D$ requires the denominator $W_{no pol}$.  
In this peculiar case it may be removed since we sort a variable that 
is absent in this denominator. 

For method 1 the implementation is more straightforward. 
Events are sorted according 
with the probability distribution eq. \ref{eq:full_pol}, subject to 
the condition \ref{eq:orthogonality}. 

In the set of figures \ref{fig:XX1}$-$\ref{fig:XX5} the distributions 
of some variables and of their correlations are reported, calculated with 
both Method 1 and 2. These show that 
Method 1 and 2 produce the same results. 

Since the full distribution shows a strong correlation between 
$\theta$ and $s_z \bar{s}_z$ on one side, and between 
$\theta$ and $s_x \bar{s}_x + s_y \bar{s}_y$ on the other side, 
the scatter plot of these pairs of variables is shown in 
figs. \ref{fig:XX1}-\ref{fig:XX2}. These figures report 
the distribution of 
100,000 events in the $(cos\theta,A)-$plane, where $A$ is 
$s_z \bar{s}_z$ (fig.\ref{fig:XX1}), 
$s_x \bar{s}_x + s_y \bar{s}_y$ (fig.\ref{fig:XX2}), 
$s_x \bar{s}_x$ (fig.\ref{fig:XX3}). 

The other figures show slices/integrals of the previous distributions. 
In fig.\ref{fig:XX4} (upper panel) I report the 
$s_z \bar{s}_z$-distribution integrated over $-0.5$ $<$ $cos(\theta)$ 
$<$ $0.5$. This region is relevant for the Collins effect that is 
suppressed for $|cos(\theta)|$ near 1. In the lower panel the 
spin-integrated $cos(\theta)-$distribution is reported. It reproduces 
the $1+cos^2(\theta)$ shape within 5 \%. 
In fig.\ref{fig:XX5} (upper panel) I report the distribution of 
$s_x \bar{s}_x+s_y \bar{s}_y$ 
integrated over $-0.5$ $<$ $cos\theta)$ 
$<$ $0.5$. In the lower panel the distribution is the one of $cos(\theta)$ 
for the integration range $-0.2$ $<$ 
$s_x \bar{s}_x+s_y \bar{s}_y$ $<$ $0.2$. In the last case, the larger 
(fluctuating) discrepancies between the distributions obtained by the 
two methods have statistical origin, and 
are due to the relatively small number of events in each bin. 

Concluding this part, I may claim that Methods 1 and 2 give similar 
results, so it is licit to apply the polarization stage of 
method 2 to events sorted by some independent generator that does 
not include spins.

\section{Step 2: Reasonable choices for the distortion methods}

Now we need to modify the transverse 
momenta of the final hadrons 
in such a way to reproduce an assigned distribution, that is 
correlated with the quark polarization. 
In an attempt to be as comprehensive as possible, I propose and test 
two classes of methods, that I name ``product'' 
and ``convolution''. These should fit the most obvious requirements. 

In this section, I apply these methods to a set of 2-momenta that 
are gaussian-distributed in the $xy-$plane with center of the 
distribution in the origin. They are supposed to be the transverse 
momenta of a set of final hadrons, originating from a parent quark 
that is directed along $+\hat z$ and has polarization $+ \hat x$. 
After testing the distorting methods on this simplified 
set, in the next section they will be applied to a set of Pythia 
events. 

Let $F_0(K_T)$ be the undistorted 
momentum distribution, that depends 
on $\vec K_T$ via $|\vec K_T|$ only. It depends on the longitudinal 
fraction $Z$ (not explicitly reported) and is proportional to the 
fragmentation function $D_1(q \rightarrow h; Z, K_T)$, where $q$ 
is one or a group of quark flavors and $h$ is one or  a group of 
hadron species. $F_0$ is normalized to 1, 
while $D_1$ is normalized to the total hadron multiplicity 
in the subset of events $q$ $\rightarrow$ $h$. 

Let $G(\vec K_T)$ be a distorting factor, that introduces 
azimuthal asymmetries in the $xy-$plane. It may depend on $Z$, 
but I do not write this explicitly. 

Product techniques: the final distribution has the 
form 
\noindent
\begin{equation}
F(\vec K_T)\ =\ F_0(K_T)\ G(\vec K_T).
\label{eq:product} 
\end{equation}

Convolution techniques: the final distribution has the 
form 
\noindent
\begin{equation}
F(\vec K_T)\ =\ \int d\vec k_T d\vec q_T 
\delta(\vec K_T\ -\ \vec k_T\ -\ \vec q_T)
\ F_0(K_T)\ G(\vec q_T).
\label{eq:convolution} 
\end{equation}
or, 
more in 
general,
\noindent
\begin{eqnarray}
F(\vec K_T)\ =\ \int d\vec k_T d\vec q_T 
\delta(\vec K_T\ -\ \vec k_T\ -\ \vec q_T)
\nonumber \\
\ F_0(K_T)\ G(\vec K_T,\vec q_T).
\label{eq:convolution2} 
\end{eqnarray}
The 
latter form is not strictly a convolution, but I will use this name 
anyway. Sometimes, 
a starting model suggests a form like in 
eq.\ref{eq:convolution}, but some constraint on 
$\vec K_T+\vec q_T$ (when sorting $\vec  q_T$) may 
remove the full independence of $G(\vec q_T)$ from $K_T$. 

The product form allows an easy implementation of parametrizations 
of the form 
\noindent
\begin{equation}
D_1(z.K_T)\ \Big(1\ +\ {{H(z,\vec K_T)} \over {D_1(z,K_T) } }\Big). 
\label{eq:product2} 
\end{equation}
The 
convolution form is more flexible to implement physical models, 
since it treats the final hadron momentum as a sum of momenta 
with independent physical origin.

\subsection{Implementation: the product case }

To implement the product form I have chosen an algorithm belonging to the 
Metropolis-Hastings family \cite{MH,MH2}: 

1) I start with an ``undistorted'' event $\vec K_T$. 

2) A shift $\vec q_T$ is sorted (flat distribution) inside a circle of 
radius $R_q$. 

3) The ``shift probability'' is 
\noindent
\begin{equation}
P(\vec K_T, \vec K_T+\vec q_T)\ \equiv\ 
{ {F(\vec K_T + \vec q_T)} \over {F(\vec K_T)} } 
\label{eq:metro1} 
\end{equation}

If this quantity is $>$ $1$, the step to the new point 
$\vec K_T+\vec q_T$ is performed. If it is not, an accept/reject 
procedure is set. As a consequence, with probability 
$P(\vec K_T, \vec K_T+\vec q_T)$ the step is performed, and 
with probability 
$1 - P(\vec K_T, \vec K_T+\vec q_T)$ the step is not performed. 

4) As a result, we have a new point $\vec K_T'$ that either coincides 
with $\vec K_T$ or with $\vec K_T+\vec q_T$. 

5) The sequence 2-4 is repeated starting with the value $\vec K_T'$ 
instead of $\vec K_T$. A new point $\vec K_T''$ is selected. 

6) The sequence 2-4 may be performed $N$ times. 

Comments: 

For $N$ $=$ 1 and for a small admitted displacement $\vec q_T$ 
the final distribution is similar to the starting one 
$F_0(K_T)$. 

For $N$ $=$ 10 the final distribution coincides with 
$F(\vec K_T)$ whichever the starting distribution was (e.g. one may choose 
$\vec K_T$ $=$ 0 fixed). So the joint choice of $N$ and $F_0(K_T)$ 
must be clever, 
in those cases where one does not know precisely the distribution of 
the undistorted events. 

\subsection{Implementation: the convolution case}

The implementation of the convolution method is more 
straightforward. 
In general terms, to get a distribution of the form 
$h(X)$ $\equiv$ 
$\int dx \int dy f(x) g(y) \delta(X-x-y)$ the steps are

1) Sort $x$ according with $f(x)$.

2) Sort $y$ according with $g(y)$.

3) Sum $x$ and $y$. 

This may work to produce both a distribution of the 
form eq. \ref{eq:convolution}, and one of the form eq. \ref{eq:convolution2}. 

\begin{figure}[ht]
\centering
\includegraphics[width=9cm]{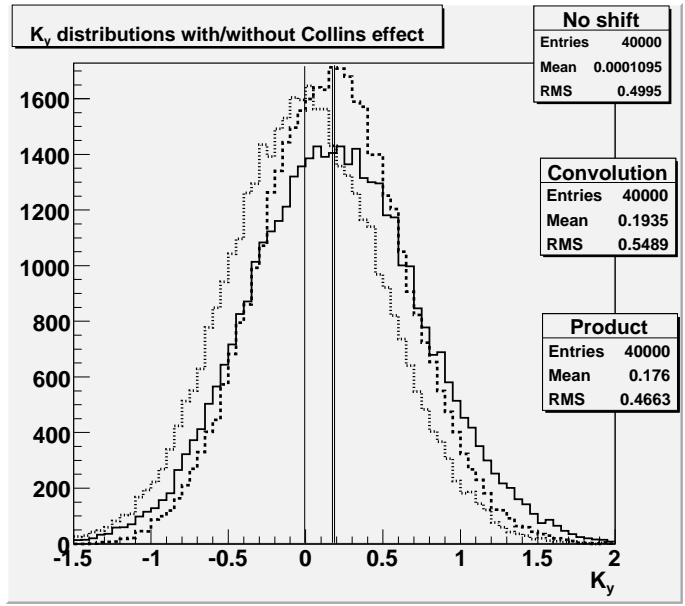}
\caption{
Distribution of 40,000 events obtained by sorting a gaussian 
distribution w.r.t. $\vec k_T$ (fine-dotted curve), 
and next by shifting events according to either the convolution method 
(continuous) or the product method (coarse-dotted). Vertical lines show the 
average point of each distribution (the two averages of the shifted 
events practically overlap). 
\label{fig:XX6}}
\end{figure}

\subsection{The examples of fig.\ref{fig:XX6}}

As an example I have chosen, both for the undistorted distribution, 
and for the distortion factors, some shapes that simplify much the 
computational work (in particular, imposing that the distorting 
factors are zero in some parts of the phase space). Apart for this, 
they do not present any special lack of generality. 

In the examples of fig.\ref{fig:XX6}, as an undistorted and axially symmetric 
distribution $F_0(K_T)$ I have chosen the 
gaussian 
\noindent
\begin{equation}
F_0(K_T)\ =\ exp[-(K_x^2+K_y^2)/(2k_1^2)] 
\label{eq:F0} 
\end{equation}
with $k_1$ $=$ 0.5 GeV/c.

For the product case I have applied eq. \ref{eq:product}
with 
\noindent
\begin{eqnarray}
G(\vec K_T)\ =\ 
\nonumber \\
\bigg(1\ +\ \alpha K_y exp[-(K_x^2+K_y^2)/(2k_2^2)]\bigg)  
\nonumber
\end{eqnarray}
when this expression is $>$ 0, 
and
\begin{eqnarray}
\\
G(\vec K_T)\ =\ 0.
\nonumber
\\
\label{eq:F1} 
\end{eqnarray}
when it is negative. 

For the example in fig.\ref{fig:XX6} 
$\alpha$ $=$ 0.75 and $k_2$ $=$ 4 GeV/c $>>$ $k_1$. At any step, the 
shift $\vec q_T$ is sorted with $q_x$ $=$ 0 and $|q_y|$ $<$ 1 GeV/c. 
I have used $N$ $=$ 10 steps, but already with 5 steps the final results 
converges towards the final required distribution. 

For the convolution case I have used 
\noindent
\begin{eqnarray}
G(\vec q_T)\ =\ 
\alpha\ q_y\ exp[-(q_x^2+q_y^2)/(2k_3^2)]  
\ for\  q_y\ >\ 0,
\nonumber
\\
G(\vec q_T)\ =\ 0\ for\ q_y\ <\ 0.
\nonumber \\
\label{eq:F2} 
\end{eqnarray}
with 
$\alpha$ $=$ 0.5 and $k_3$ $=$ 0.1 GeV/c. These parameters are chosen so 
that the convolution produces a shift that is similar to the product 
case. 
The choices $k_2$ 
$>>$ $k_1$ (product case), and $k_3$ $<<$ $k_1$ (convolution case) 
lead to a similar result, i.e. avoiding  
that the final distribution is much broader than the starting one. 
This permits us to appreciate the shifting effect, that in 
fig. \ref{fig:XX6} is highlighted by vertical lines showing the 
average point of each distribution. 

\section{Example: A $cos(\Phi_1+\Phi_2)-$asymmetry in the complete event}

As an example, the two described track-distortion methods, with 
the same parameters as in fig.\ref{fig:XX6} are applied to produce a 
$cos(\Phi_1+\Phi_2)$ nonzero term in the distribution of pion pairs 
generated by Pythia in the conditions of the Belle experiment 
in the sub-$b\bar{b}$ energy range, at $Q$ $=$ 
10.0 GeV (Belle has performed measurements in this range, and 
at a slightly higher 
$Q$ at the $b\bar{b}$ threshold\cite{BelleCollins2}). 

To apply the previous methods to 
events produced by Pythia some further complications are 
needed. 
I do not give details on these points since they just require standard 
operations like frame rotations and vector projections. 
What is done is (1) the distortion of the hadron transverse  momentum 
is produced in a frame where the parent quark (or 
antiquark) momentum and spin are along 
the $z$ and $x$ axes (this reproduces the situation analyzed in the previous 
section), (2) for the specific data analysis (i.e. extraction of the
$cos(\Phi_1+\Phi_2)-$distribution) each resulting hadron momentum 
is transferred to a frame where the lepton-quark scattering 
plane coincides with the a coordinate plane. 

The Pythia-generated events are selected by 
the additional condition Thrust $>$ 0.8. This 
excludes three-jet events, and in the case of $Q$ slightly over 
10 GeV (i.e. over the value used here) 
it would exclude events of $b\bar{b}-$kind (see the discussion 
in \cite{BelleCollins2}). This cutoff is important, because the 
momentum-spin correlation calculated in the Appendix of the present 
work refers to ``light'' quarks, i.e. fermions for which it is 
licit to assume helicity conservation in the vector-fermion vertex. 
At $Q$ $=$ 10 GeV we have to include 
events starting from $u\bar{u}$, $d\bar{d}$ $s\bar{s}$ 
and $c\bar{c}$ pairs. In the collision c.m. frame, 
the quark energy is 5 GeV. The charm mass is $\approx$ 1.27 
GeV/c$^2$. For this quark the first order correction to the 
UR relation $E$ $=$ $P$ is $m^2/2E$ $\approx$ 160 MeV $<<$ $E$, 
so the helicity nonconserving terms have small relevance.  
If gluon radiation processes with a large $p_T$ were included 
in the hard photon-quark vertex, the spin-momentum correlations 
would not be exactly as in eq. \ref{eq:full_pol} (e.g. the quark 
momentum would be modified by a non-collinear gluon radiation, 
and the large quark virtuality would affect 
helicity conservation in the photon vertex). 
All these processes are excluded by the request Thrust $>$ 0.8. 

According with e.g. \cite{BoerBelle1} in presence of nonzero 
Collins functions on both sides one expects a cross section of the 
form 
\begin{eqnarray}
{{d \sigma} \over {dz_1dz_2 d\Omega_1d\Omega_2}}\ \propto\
\Big[1\ +\ cos^2(\theta)\Big]D(z_1)D(z_2)\ 
\nonumber \\ 
+\ sin(\theta) cos(\Phi_1+\Phi_2) C(z_1) C(z_2).
\label{eq:Boer1}
\end{eqnarray}
The
angle $\theta$ is the lepton-quark polar angle. 
$\Phi_{1,2}$ are the angles of the pion transverse momenta 
w.r.t. the lepton-quark scattering plane. $D(z_{1,2})$ and 
$C(z_{1,2})$ are the unpolarized and Collins fragmentation functions 
with opposite hemisphere pions with longitudinal fractions $z_1$ and 
$z_2$. 

Since this is just an example of  
application, I have not attempted reproducing the recently 
measured values for this quantity at 
Belle\cite{BelleCollins1,BelleCollins2}. Rather, I have reported 
the output corresponding to the distribution shifts reported 
in fig.\ref{fig:XX6}, including the ``no-shift'' case to have an 
estimator of the fake asymmetries due to the statistics and to the 
cutoffs. 

\begin{figure}[ht]
\centering
\includegraphics[width=9cm]{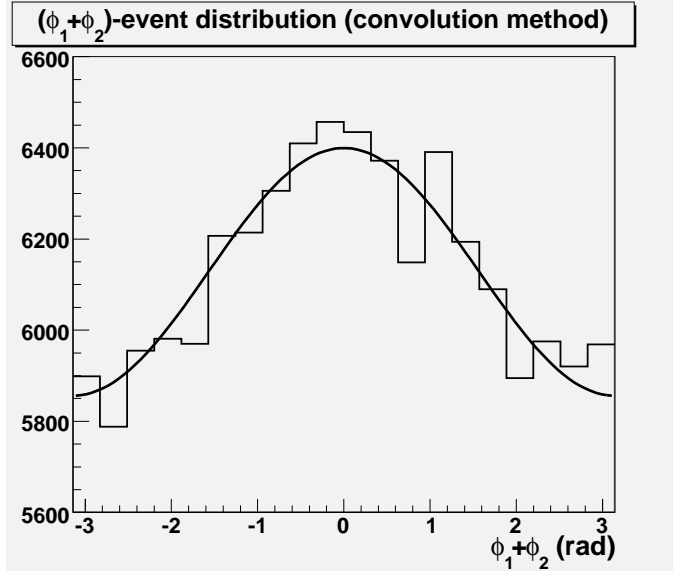}
\caption{
Distribution of the angle $\Phi_1+\Phi_2$, where the angles 
$\Phi_i$ are the azimuthal angles (see text) of two 
opposite-hemisphere pions. The individual 
pion tracks are deflected according with the convolution method. 
The fitting curve is $\propto$ $1 + A\ cos(\Phi_1+\Phi_2)$, with 
$A$ $=$ 0.04 $\pm$ 0.004. 
\label{fig:XX7}}
\end{figure}

\begin{figure}[ht]
\centering
\includegraphics[width=9cm]{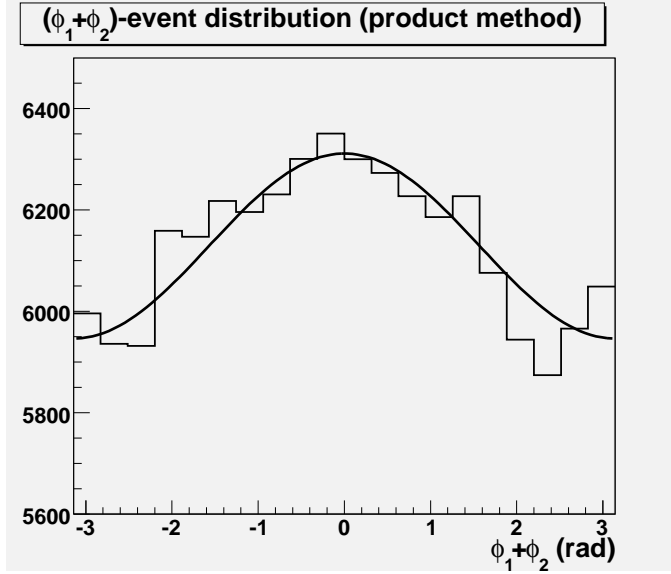}
\caption{
Distribution of the angle $\Phi_1+\Phi_2$, as in fig.\ref{fig:XX7}. 
The individual pion tracks are here deflected according with 
the product method. 
The fitting curve is $\propto$ $1 + A\ cos(\Phi_1+\Phi_2)$, with 
$A$ $=$ 0.03 $\pm$ 0.004. 
\label{fig:XX8}}
\end{figure}

\begin{figure}[ht]
\centering
\includegraphics[width=9cm]{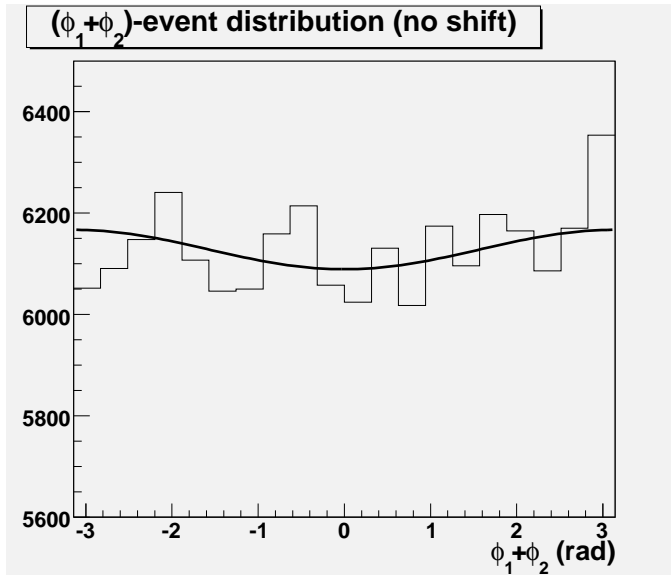}
\caption{
Distribution of the angle $\Phi_1+\Phi_2$, as in fig.\ref{fig:XX7}. 
The individual pion tracks are not deflected at all, i.e. they are the 
original trajectories as simulated by Pythia. 
The fitting curve is $\propto$ $1 + A\ cos(\Phi_1+\Phi_2)$, with 
$A$ $=$ $-$0.006 $\pm$ 0.004. 
\label{fig:XX9}}
\end{figure}

The three figures \ref{fig:XX7}, \ref{fig:XX8}, \ref{fig:XX9} 
show the distribution of 135,000 pion pairs vs $cos(\Phi_1+\Phi_2)$ 
in Belle conditions. Each event has Thrust $>$ 0.8, and each 
individual pion has $z$ $>$ 0.1, and $cos(\alpha$ $>$ 0.8, where 
$\alpha$ is the angle w.r.t. the quark axis (in an attempt of 
precisely fitting 
the experimental data, one should include further cuts and use 
the Thrust axis instead of the quark axis). A ``pair'' is composed 
of two pions (regardless of their charge) belonging to different 
hemispheres. 

The produced histograms have been fitted by the Root-Migrad 
package\cite{Root} 
with curves of the form 
$B [1 + A\ cos(\Phi_1+\Phi_2)]$. The data are divided into 40 equal-range 
bins, and $B$ $=$ 6130 $\pm$ 20 is the average number of events per bin. 
$A$ has 
size 3-4 \%, compared to a statistical error 0.4 \% and to a 
zero-asymmetry value 0.6 \% (extracted from  
the last figure, where no azimuthal effect is included).  

The chosen form of the spin-dependent fragmentation functions 
is aimed at simplifying the implementation of the presented 
examples. I have 
not considered a dependence of the azimuthal effects on the longitudinal 
fractions. 
I have not considered the flavor dependence of the spin-dependent 
fragmentation functions. I 
have only considered spin effects on pions. 
To include different functional forms, a dependence on $z$, flavor 
differentiation, azimuthal effects on other hadron species 
(like kaons), would be a matter of more program lines, but 
would not issue any further challenge. 

The same procedure could have been applied to an 
intermediate state hadron that decayed into two hadrons like e.g. 
a $\rho$. Then the two decay hadrons would present an extra 
fragmentation function of the kind $H_\perp$ (see 
\cite{BBJR1}, \cite{BBJR2}).  
On the other side, to produce a function of the kind $H^<$ 
(\cite{BBJR1}, \cite{BBJR2}, in $e^+e^-$ $\rightarrow$ $hadrons$
it has been measured by \cite{Vossen}) modifications 
should be applied 
to the $relative$ momentum of the two final hadrons. These 
are straightforward generalizations of the work presented here. 
Another possible application is the production of polarized 
$\Lambda/\bar{\Lambda}$ \cite{Lambda}. In this case one would 
sort a spin for this hadron correlated with the quark spin, and 
sort new momenta for the decay products according with 
the assigned spin. 

\section{Conclusions}

I have started this work from 
events of the class $e^+e^-$ $\rightarrow$ $hadrons$, generated by an  
ordinary Monte Carlo code where the transverse polarizations of the quarks 
are not taken into account. 

A set of techniques have been presented to build code patches that do not 
touch the Monte Carlo generator itself, but rather modify its outputs, 
both at parton and hadron level. 

The first aim of these modifications is attributing a transverse spin to 
the quark and to the antiquark pair produced in the hard vertex, in such 
a way to get a physically sound distribution for the correlated set of 
quark momenta and spins produced in the hard vertex. 
To reach this, $e^+e^-$ events with production of polarized quarks have 
been generated according to a known and general method. Next, 
an alternative two-step method for generating similar events 
has been tested, where in the first 
step an unpolarized quark and an unpolarized antiquark are produced 
in the lepton annihilation, 
and in the second step both are polarized without touching 
the previously sorted momenta. I have shown that the 
two methods produce the same distributions. This means that it is 
licit to take an unpolarized quark-antiquark pair produced from 
an external generator and attribute a pair of polarizations to it 
according to the second step of the two-step method. 

The second aim is to slightly modify the transverse momenta of the 
produced hadrons, in a way that is related to the spin of the parent  
quark or antiquark, and controllable. ``Controllable'' may mean two 
alternative possibilities: 
Either that a given physical model is implemented, or that we know 
the form of the momentum distribution, or of the fragmentation function, 
that we want to obtain. 
Two classes of techniques have been presented, namely
``product'' and ``convolution'' techniques. The former group is 
suitable for 
producing final distributions according with fragmentation functions 
of the form $D(1+H/D)$, where $D$ is the unpolarized fragmentation 
function and $H$ a spin-dependent term, like e.g. a Collins function.  
The ``convolution'' techniques are more appropriate for those cases where 
a model predicts that the final moment is a convolution of two 
contributions, one due to the unpolarized physics, and the other one 
to polarization-related effects. 

As a test case, this has been applied to modification of a set of Pythia 
events, so to produce an azimuthal asymmetry that derives from the 
combined effect 
of the Collins functions of two opposite-produced pions. This was just a 
test case, the range of possible applications is quite large. 

As a final and due observation, I remark that the events produced  
by the methods described here, working on Pythia outputs,   
are $not$ Pythia 
events. They are modifications of Pythia events.

$\\$

{\bf Acknowledgments}
$\\$

This work is a ``side effect'' of a long series of thorough discussions with 
Alessandro Bacchetta, on the physics of the fragmentation functions 
and on the tools to analyze them.

\section{Appendix: Polarized quark $-$ unpolarized lepton contraction 
$H^{\mu\nu}(s_y,\bar{s_y})L_{\mu\nu}$}

For the procedure described in section 2 we need the contraction 
\noindent
\begin{equation}
q^4 W\ \equiv\ H^{\mu\nu}L_{\mu\nu}
\label{eq:W} 
\end{equation}
of the quark-level hadronic and lepton tensors 
in the processes 
$l^+l^-$ $\rightarrow$ $q\bar{q}$, where the final pair is constituted by 
a transverse-polarized quark and antiquark. This will be calculated for purely 
massless quarks and leptons, in the center of mass frame of the reaction. 

Eq.\ref{eq:W} gives, for assigned momenta of the electrons, 
the joint probability for sorting the momenta and the spins of the quark 
and the antiquark. In a lepton annihilation process, $q$ has no 
probabilistic role since it is fixed. 

In ref. \cite{AB_EPJA2} (see the Appendix of that work) 
the $q^4W$ factor was already calculated for 
the reversed process 
$q\bar{q}$ $\rightarrow$ $l^+l^-$, aimed at the Drell-Yan application. 
The invariant result is the same in both 
cases, however relevant differences appear when one rewrites it  
in the reference frame where the transverse spins need to be 
generated. 
The present case is simpler than the Drell-Yan one and requires 
no approximations, since the leptons are always on the 
$z-$axis, and the polarized 
quark and antiquark are exactly back-to-back. 
Because of this, the transverse spins of the quark and of the antiquark 
are orthogonal to the same axis. The 4-vector 
associated to the quark spin is orthogonal to the 4-momenta of the quark 
and of the antiquark. The former orthogonality is always true, the latter 
only for back-to-back pairs. 

I 
use the shortened  notation for 
traces 
\begin{equation}
T[..]\ \equiv\ {1 \over 4} Tr[..].  
\end{equation}
I use the definitions of the Berestevskij-Lifsits-Pitaevskij  
book \cite{LL4} (better known as the 4th book of the Landau-Lifsitz 
Course in Theoretical Physics). As the only exception to this, I 
indicate $k_\mu\gamma^\mu$ with the widespread notation $\ksl$ 
instead of using $\hat k$ as was done in that book. 

\subsection{The case of unpolarized quarks}

If 
the quarks are unpolarized, we simply 
have 
\noindent
\begin{eqnarray}
H_{unpol}^{\mu\nu}\ \equiv\ T[\ksl\gmu\kbsl\gnu]\ 
=\ \{\kmu,\kbnu\}\ -\ (k\cdot \bar{k}) \gmunu
\nonumber\\
=\ \{\kmu,\kbnu\}\ -\ {q^2 \over 2} \gmunu
\label{eq:hunpol} 
\end{eqnarray}
and 

\noindent
\begin{equation}
L^{\mu\nu}\ \equiv\ 
\{\pmu,\pbnu\}\ -\ {q^2 \over 2} \gmunu
\label{eq:l0} 
\end{equation}
where 
braces indicate symmetric dyadic product. 
The contraction of the two is faster 
using 
\noindent
\begin{equation}
\{k_\mu,\bar{k}_\nu\} \{\pmu,\pbnu\}\ =\ 
2\ [(kp)(\bar{k}\bar{p})\ +\ (k\bar{p})(\bar{k}p)],
\label{eq:aus1} 
\end{equation}

\noindent
\begin{equation}
\Big(
\{k_\mu,\bar{k}_\nu\}\ -\ (k\bar{k}) g_{\mu\nu}
\Big)\ \gmunu\ =\ 0. 
\label{eq:aus2} 
\end{equation}

\noindent
We get
\noindent
\begin{eqnarray}
q^4 W_{unpol}\ \equiv\ T[\ksl\gmu\kbsl\gnu]\ T[\psl\gamma_\mu\pbsl\gamma_\nu]\ 
\ =\ 
\nonumber\\
2\ [(kp)(\bar{k}\bar{p})\ +\ (k\bar{p})(\bar{k}p)].
\label{eq:H0} 
\end{eqnarray}

I 
extract a factor $q$ from each vector:

\noindent
\begin{equation}
\kmu\ \equiv\ (q/2) \Nmu, \ \pmu\ \equiv\ (q/2) \nmu, ...
\label{eq:redef1} 
\end{equation}
In a center of mass frame of the partonic process, the 3-vectors 
$\vec N$, $\vec n$ etc are unitary vectors. 
I get
\noindent
\begin{equation}
W_{unpol}\ 
\ =\ 
{1 \over 8}\ [(Nn)(\bar{N}\bar{n})\ +\ (N\bar{n})(\bar{N}n)].
\label{eq:H0b} 
\end{equation}
A more familiar way to write this may be 
\noindent
\begin{equation}
W_{unpol}\ 
\ =\ 
{1 \over 4}\ [1\ +\ cos^2(\theta)].
\label{eq:H0c} 
\end{equation}
where 
$\theta$ is the angle between the lepton and quark directions in 
the partonic center of mass frame. 

\subsection{polarized quarks}

In the case of polarized (anti)quark, we need to substitute 

\noindent
\begin{equation}
\ksl\ \rightarrow\ \ksl(1-\g^5\ssl)
\label{eq:density} 
\end{equation}
where
$s^\mu$ is the polarization 4-vector for the quark, respecting 
the exact 4-dimensional constraint $(sk)$ $=$ 0. 
If each spin is exactly 
transverse  to the corresponding momentum, then also $\vec s \cdot \vec k$ 
$=$ 0. 

The relation between $s^\mu$ and the polarization in a 
rest frame 
$\vec \sigma$ is, for 
massive 
particles,  
\noindent
\begin{equation}
s_0\ =\ \sigma_L |\vec k|/m,\ s_L\ =\ \sigma_L E/m,\ 
\vec s_T\ =\ \vec \sigma_T
\label{eq:spin1} 
\end{equation}
and 
evidently it creates problems for $m$ $\approx$ 0, unless the longitudinal 
component is strictly zero. 
However, when the previous expressions are used to write 
the density matrix eq.\ref{eq:density} in terms of the 
rest frame polarizations, $E/m-$terms cancel and we may check that 
the density matrix is free from mass 
singularities: 
\noindent
\begin{equation}
\ksl(1-\g^5\ssl)\ =\ 
\ksl [1 - \g^5 (\pm \sigma_L + \vec \sigma_T \cdot \gamma_T)].
\label{eq:spin2b} 
\end{equation} 
($\pm$ differentiates particles and antiparticles).  

Although it is not the aim of this particular work, it is useful to consider 
what would be the outcome for longitudinally polarized quarks, with helicities 
$h$ and $\bar{h}$: 

\noindent
\begin{eqnarray}
W_{unpol}\ \rightarrow\ W_{h,\bar{h}}\ =\ 
\nonumber\\
(1-h\bar{h}) W_{unpol} 
\ \propto\ (1-h\bar{h}) [1\ +\ cos^2(\theta)].
\label{eq:long} 
\end{eqnarray}
This 
is a predictable and known result for massless fermions: 
pairs with the same helicity are suppressed 
since they have opposite longitudinal spins in the c.m., i.e. total 
longitudinal spin zero. This cannot be transferred immediately to the 
transverse spin, since any helicity component is composed by 50 \% 
$\pm$ components along any chosen transverse axis. In addition, 
the $1+cos^2(\theta)$ factor is a composition of $Y_{00}$ and 
$Y_{10}$ eigenfunctions of the orbital angular momentum along the $z-$axis, 
not along a transverse axis. So, while eq.\ref{eq:long} could 
be guessed from the very beginning, a statement like ``the transverse 
spins of the pair are mostly parallel'' has not a solid $a$ $priori$ 
justification. 

The full trace for polarized quarks is 
\noindent
\begin{equation}
H_{s,\bar{s}}^{\mu\nu}\ \equiv\ 
T[\ksl\gmu(1 - \g^5\ssl)\kbsl\gnu(1 - \g^5\sbsl)]
\label{eq:full1} 
\end{equation}
that excluding 
$O\bigg((\g^5)^1\bigg)$ terms 
(these would lead to an antisymmetric tensor, that is useless when 
contracted with the symmetric tensor of the unpolarized leptons) 
reduces to 
\noindent
\begin{equation}
H_{s,\bar{s}}^{\mu\nu}\ \equiv\ 
T[\ksl\gmu\kbsl\gnu]\ +\ 
T[\ksl\gmu\g^5\ssl\kbsl\gnu\g^5\sbsl]
\label{eq:full2} 
\end{equation}
\noindent
\begin{equation}
=\ 
T[\ksl\gmu\kbsl\gnu]\ -\ 
T[\ksl\gmu\ssl\kbsl\gnu\sbsl].
\label{eq:full3} 
\end{equation}

Next I use the reduction formula
\noindent
\begin{eqnarray}
T(abcdef)\ =\ g^{ab}T(cdef) - g^{ac}T(bdef) + 
\nonumber
\\
+\ g^{ad}T(bcef) - 
g^{ae}T(bcdf) + 
g^{af}T(bcde) 
\label{eq:reduce1} 
\end{eqnarray}
(with 
the obvious notation $T(ab...)$ $\equiv$ $T[\gamma^a\gamma^b...]$). 
For 
any of the remaining traces we have the better known 
relation 
\noindent
\begin{equation}
T(abcd)\ =\ g^{ab}g^{cd} - g^{ac}g^{bd} + g^{ad}g^{bc}
\label{eq:reduce2} 
\end{equation}
The composition of eqs. \ref{eq:reduce1} and \ref{eq:reduce2} 
produces 15 terms. Applied to the second term of eq. \ref{eq:full3}, 
two of these terms contain the products $(sk) = 0$ and 
$(\bar{s}\bar{k}) = 0$ and are dropped. 

In the specific case of the transverse spins for a back-to-back 
$q\bar{q}$ pair we also have 
$(\bar{s}k) = 0$ and $(s\bar{k}) = 0$. So most terms are dropped and 
we are left 
with 
\noindent
\begin{eqnarray}
\Bigg(
[1 - (s\bar{s})]\cdot H^{\mu\nu}_{unpol}\ -\ 
{q^2 \over 2} \{\smu,\snu\} \Bigg)
\label{eq:hfull1} 
\end{eqnarray}

After contracting the hadron tensor with the (unpolarized) lepton tensor, 
the final result is 

\noindent
\begin{eqnarray}
W_{s_T\bar{s_T}}\ =\ 
{1 \over 4}
\Bigg(
[1\ +\ cos^2(\theta)] (1\ +\ s_z\bar{s}_z)\ 
\nonumber\\
-\ sin^2(\theta) (s_x\bar{s}_x\ +\ s_y\bar{s}_y) 
\Bigg)
\label{eq:full2b} 
\end{eqnarray}
Here
I have used eq.\ref{eq:H0c}. The appearance of nonzero 
$s_z$, $\bar{s}_z$ terms should not confuse about the transverse 
nature of the considered spins. These terms appear 
because the $z-$axis is parallel to the lepton momentum, not to 
the quark momentum. 

The $s_i\bar{s}_i-$terms may have two different origins in the 
previous invariant equations: from terms like 
$(s\bar{s})$, 
and from terms like $(sp)(\bar{s}\bar{p})$. When passing to 
3-dimensional components, both terms change sign. The former because 
$(s\bar{s})$ $\rightarrow$ $-(\vec s \cdot \vec {\bar{s}})$. 
The latter term presents two such changes of sign, and a third due 
to the opposite space parts of $p_\mu$ and $\bar{p}_\mu$.




\end{document}